\documentclass[epjST_arXiv]{svjour_arXiv}
\usepackage{amsmath}
\usepackage{graphicx,amssymb,amsfonts,latexsym,color,dcolumn,bm, braket,dsfont}
\usepackage{soul}

\newcommand{\HH}{\mathcal{H}}
\newcommand{\CC}{\mathcal{C}}
\renewcommand{\imath}{\mathrm{i}}

\begin{document}

\title{Homodyne versus photon-counting quantum trajectories for dissipative Kerr resonators with two-photon driving}

\author{
Nicola Bartolo\thanks{\email{nicola.bartolo@univ-paris-diderot.fr}}
\and
Fabrizio Minganti\thanks{The first two authors contributed equally to this work.}
\and
Jared Lolli
\and
Cristiano Ciuti\thanks{\email{cristiano.ciuti@univ-paris-diderot.fr}}}
\institute{Universit\'e Paris Diderot, Sorbonne Paris Cit\'e, Laboratoire Mat\'eriaux et Ph\'enom\`enes Quantiques, CNRS-UMR7162, 75013 Paris, France}

\abstract{
We investigate two different kinds of quantum trajectories for a nonlinear photon resonator subject to two-photon pumping, a configuration recently studied for the generation of photonic Schr\"odinger cat states. 
In the absence of feedback control and in the strong-driving limit, the steady-state density matrix is a statistical mixture of two states with equal weight.
While along a single photon-counting trajectory the systems intermittently switches between an odd and an even cat state, we show that upon homodyne detection the situation is different.
Indeed, homodyne quantum trajectories reveal switches between coherent states of opposite phase.}

\maketitle

\section{Introduction}\label{Sec:Intro}

In the last years, a growing interest has been shown towards the physics of strongly-correlated systems, characterised by strong interactions between their composing particles.
Several platforms are nowadays available to experimentally access this condition.
Among them, cold atoms in optical lattices~\cite{BlochNatPhys03,BlochRMP08} and trapped ions~\cite{LeibfriedRMP03,BlattNatPhys12} represent paradigmatic tools of investigation, allowing a fine control on the interactions and a high-resolution imaging of the system.
Recently, strongly-interacting photons in semiconductor microcavities~\cite{WeisbuchPRL92,DeveaudBOOK} and superconducting circuits~\cite{SchoelkopfNature08,YouNature11} opened new perspectives in this field.
Indeed, the light-matter interaction in nonlinear materials can mediate an effective non-negligible photon-photon interaction~(cf. \cite{CarusottoRMP13} and references therein).
Since an ensemble of photons is a bosonic quantum gas, the addition of interaction can produce many-body effects.
Differently from trapped gases and ions, the leak of photons from a resonator can not be neglected on experimental timescales, and therefore photons must be continuously pumped into the system. 
The competition between drive, dissipation, and interactions in such kind of out-of-equilibrium quantum systems enriches the physical scenario.

In this work, we investigate the properties of a Kerr resonator of frequency $\omega_c$ subject to engineered two-photon pumping and dissipation \cite{LeghtasScience15}.
The steady state of this model has a peculiar bimodal character, which is connected to the emergence of photonic Schr\"odinger's cat states \cite{MingantiSciRep16}.
To shed light on the elusive features of quantum bimodality,
we analyse this system via the quantum trajectory method~\cite{DalibardPRL92,CarmichaelPRL93,MolmerJOSAB93,PlenioRMP98}.

\section{The system}

The system under consideration is a single nonlinear Kerr resonator subject to a parametric two-photon driving of frequency $2\omega_c$.
The time dynamics of the density matrix $\hat{\rho}$ is ruled by the Lindblad super operator $\mathcal{L}$ via the master equation $i \partial_t \hat{\rho} = \mathcal{L} \hat{\rho}$.
The superoperator $\mathcal{L}$ includes a Hamiltonian evolution and non-hermitian contributions which describe the dissipation processes, as detailed in e.g.~\cite{HarocheBOOK,CarmichaelBOOK,WallsBOOK}.
In a frame rotating at $\omega_c$, the Hamiltonian reads
\begin{equation}\label{Eq:Hamiltonian}
\hat{\HH}
=\frac{U}{2}\,\hat{a}^\dagger\hat{a}^\dagger\hat{a}\hat{a}
+\frac{G}{2}\left(\hat{a}^\dagger\hat{a}^\dagger+\hat{a}\hat{a}\right).
\end{equation}
In the equation above, $U$ is the Kerr photon-photon interaction strength, $G$ is the two-photon driving amplitude, and $\hat{a}^\dagger$ ($\hat{a}$) is the creation (annihilation) operator of the photonic field.
If we include one- and two-photon dissipation processes, the Lindblad superoperator acts according to $(\hbar=1)$
\begin{equation}\label{Eq:Lindblad}
\mathcal{L} \hat{\rho} = \left[\hat{\HH},\hat{\rho}\right]
+\imath\, \frac{\gamma}{2} \left(2\hat{a}\hat{\rho}\hat{a}^\dagger
-\hat{a}^\dagger\hat{a}\hat{\rho}
-\hat{\rho}\hat{a}^\dagger\hat{a}\right)
+\imath\, \frac{\eta}{2} \left(2\hat{a}\hat{a}\hat{\rho}\hat{a}^\dagger\hat{a}^\dagger
-\hat{a}^\dagger\hat{a}^\dagger\hat{a}\hat{a}\hat{\rho}
-\hat{\rho}\hat{a}^\dagger\hat{a}^\dagger\hat{a}\hat{a}\right),
\end{equation}
where $\gamma$ and $\eta$ are, respectively, the one- and two-photon dissipation rates.
We point out that two-photon drive and dissipation have been already implemented via reservoir engineering~\cite{LeghtasScience15}.
While, as usual, the one-photon losses are due to the finite quality factor of the resonator, two-photon losses are a consequence of the engineered driving.

The model described by the master equation~\eqref{Eq:Lindblad} can be solved exactly for its steady state~\cite{MingantiSciRep16,KryuchkyanOC96,MeaneyEPJQT14,ElliottPRA16,BartoloPRA16}.
For a broad range of parameters, the corresponding density matrix $\hat{\rho}_{\rm ss}$ is well approximated by the statistical mixture of two orthogonal states:
\begin{equation}\label{Eq:MixtureCats}
\hat{\rho}_{\rm ss}\simeq
p^+\,\ket{\CC^+_\alpha}\!\bra{\CC^+_\alpha}
+p^-\,\ket{\CC^-_\alpha}\!\bra{\CC^-_\alpha},
\end{equation}
where $\ket{\CC^\pm_\alpha}\propto\ket{\alpha}\pm\ket{-\alpha}$ are photonic Schr\"odinger cat states whose complex amplitude $\alpha$ is determined by the system parameters~\cite{LeghtasScience15,MingantiSciRep16}.
We recall that the coherent state $\ket{\alpha}$ is the eigenstate of the destruction operator: $\hat{a} \ket{\alpha}=\alpha \ket{\alpha}$.
The state $\ket{\CC^+_\alpha}$ is called the even cat, since it can be written as a superposition of solely even Fock states, while $\ket{\CC^-_\alpha}$ is the odd cat. 
In Eq.\eqref{Eq:MixtureCats}, the coefficients $p^\pm$ can be interpreted as the probabilities of the system of being found in the corresponding cat state.
For intense pumping ($G\gg U,\gamma,\eta$), one has $|\alpha|\gg1$ and $p^+\simeq p^- \simeq 1/2$.
However, in this strong-pumping regime, Eq.~\eqref{Eq:MixtureCats} can be recast as
\begin{equation}\label{Eq:MixtureCoherent}
\hat{\rho}_{\rm ss}\simeq
\frac{1}{2}\ket{\alpha}\!\bra{\alpha}
+\frac{1}{2}\ket{-\alpha}\!\bra{-\alpha}.
\end{equation}
Hence, the steady state can be seen as well as a statistical mixture of two coherent states of opposite phase.
Since $\hat{\rho}_{\rm ss}$ is anyhow a mixture of two (quasi-)orthogonal states, the steady state is bimodal. 
Such a bimodality can be visualised, for instance, through the Wigner function \cite{MingantiSciRep16,BartoloPRA16}.
Now, the pivotal question is:  if one monitors the evolution of the system, in which states can it be observed?
The orthogonal cat states in Eq.~\eqref{Eq:MixtureCats} with $p^\pm= 1/2$, the two coherent states with opposite phases in Eq.~\eqref{Eq:MixtureCoherent}, or none of them in particular?
As we will show in the following, the answer dramatically depends on the type of measurement scheme employed to monitor the trajectory of the system.

\section{The quantum trajectory approach}

From a theoretical point of view, the Lindblad master equation describes the out-of-equilibrium dynamics of a system coupled to a Markovian (i.e., memoryless) environment.
Indeed, the density matrix $\hat{\rho}(t)$ obtained by solving Eq.~\eqref{Eq:Lindblad} encodes the average evolution of the system when no information is collected about the environment state.
	On the other hand, one can imagine to keep track of the system state by continuously probing the environment.
	Doing so, the time evolution of the system would change at each realisation.
	However, $\hat{\rho}(t)$ could be retrieved by averaging over an infinite number of such ``monitored'' realisations.

The Montecarlo wavefunction method has been developed relying exactly on this idea.
	It is based on the stochastic simulation of the system evolution when one continuously gathers information from the environment.
Each simulation of the stochastic evolution of the system gives a single quantum trajectory.
The results obtained by solving the master equation~\eqref{Eq:Lindblad} are recovered by averaging over many trajectories.
In order to simulate the quantum trajectories, it is necessary  to explicitly model how an observer measures the environment, thus affecting the system evolution itself (a detailed  discussion on this subject is given in \cite{HarocheBOOK,CarmichaelBOOK}).
Interestingly, several different measures can be associated with the same master equation.
Depending on the chosen measurement, contrasting results and interpretations can emerge.
Those incompatibilities are, however, harmonized once the mean value over many trajectories is taken.
In the following, we briefly introduce the quantum trajectory formalism for photon counting and for homodyne detection, both associated to the same master equation~\eqref{Eq:Lindblad}.

\subsection{Photon counting}

The most natural way to observe the exchanges between the Kerr resonator and the environment is to just detect every leaked photon (both individually and by couples).
This mechanism is described via the action of the one-photon jump operator $\hat{J}_1=\sqrt{\gamma}\, \hat{a}$ and the two-photon one $\hat{J}_2=\sqrt{\eta}\, \hat{a}^2$, which describe the absorption of one or two photons by an ideal photodetector (details in e.g.~\cite{WisemanBOOK}).
Indeed, in typical realisations (e.g. \cite{LeghtasScience15}) the one- and two-photon dissipation channels are discernible.
Hence, we can assume that the photodetector is capable of distinguishing between one- and two-photon losses.
The photon-counting trajectory is then obtained by discretising the system time evolution.
At each time step, one stochastically determines if a single photon or a couple of them has been detected.
To do so, one considers that the probability of a one- and two-photon detection in a time step $dt$ are, respectively,
\begin{equation}
P_1(dt)= \langle \hat{J}^\dagger_1 \hat{J}_1\rangle dt=\gamma \langle \hat{a}^\dagger \hat{a} \rangle dt, \qquad  P_2(dt)=\langle \hat{J}^\dagger_2 \hat{J}_2\rangle dt=\eta \langle \hat{a}^{\dagger\, 2} \hat{a}^2 \rangle dt.
\end{equation}
If a jump occurs, the system state abruptly changes under the action of the corresponding jump operator according to
\begin{equation}
\ket{\Psi(t+dt)}\propto \hat{J}_{\nu}\ket{\Psi(t)},\qquad\nu=1,2,
\end{equation}
upon appropriate normalization of the wave function.
If no jump occurs, the state evolves under the action of an effective non-hermitian Hamiltonian operator:
\begin{equation}\label{Eq:PhotonCounting}
\frac{d \ket{\Psi(t)}}{dt}= - \imath \left(  \hat{\HH}
-  \frac{\imath}{2} \sum_{\nu=1,2} \hat{J}^\dagger_\nu \hat{J}_\nu \right) \ket{\Psi(t)}.
\end{equation}
We stress that $dt$ must be sufficiently small to ensure: (i) $P_{1,2}(dt) \ll 1$, such to avoid multiple jumps in the same time step; (ii) a smooth numerical integration of Eq.~\eqref{Eq:PhotonCounting}.
In conclusion, a photon-counting trajectory is characterised by abrupt jumps corresponding to the projective measure associated to the detection of one or two photons.

\subsection{Homodyne detection}

Another possible way to monitor a quantum-optical system is through homodyne detection, a widely-used experimental technique which allows to access the field quadratures~\cite{SmitheyPRL93,ZavattaPRA04,CampagnePRX16}.
To implement this kind of measurement, the cavity output field is mixed to the coherent field of a reference laser through a beam splitter (here assumed of perfect transmittance).
Then, the mixed fields are probed via (perfect) photodetectors, whose measures are described by new jump operators.
We stress that both the coherent and the cavity fields are measured simultaneously.

In our case, we want to probe independently the two dissipation channels.
To distinguish between one- and two-photon losses, one can exploit a nonlinear element acting on the cavity output field.
Indeed, in experimental realisations such as~\cite{LeghtasScience15}, a nonlinear element is already part of the system and is the key ingredient to realise two-photon processes.
More specifically, one-photon losses are due to the finite quality factor of the resonator.
They can be probed by directly mixing the output field of the cavity with a coherent beam of amplitude $\beta_1$ acting as local oscillator.
Therefore, the homodyne jump operator for one-photon losses can be cast as $\hat{K}_1=\hat{J}_1 +\beta_1 \hat{\mathds{1}}$.
Two-photon losses are, instead, mediated by a nonlinear element (a Josephson junction in~\cite{LeghtasScience15}), which converts two cavity photons of frequency $\omega_c$ into one photon of frequency $\omega_{nl}$. Hence, the field coming out of the nonlinear element can be probed by a second independent oscillator.
This whole process can be seen as the action of a nonlinear beam splitter which mixes couples of dissipated photons with a reference oscillator of amplitude $\beta_2$.
Therefore, the homodyne two-photon jump operator takes the form $\hat{K}_2=\hat{J}_2 +\beta_2 \hat{\mathds{1}}$.
Without loss of generality, in the following, we assume the amplitudes $\beta_{1,2}$ to be real~\cite{WisemanBOOK}.

From the definitions of the jump operators, one extracts the jump probabilities
\begin{align}
P_1(dt)&= \langle \hat{K}^\dagger_1 \hat{K}_1\rangle dt=\langle (\sqrt{\gamma}\hat{a} + \beta_1\hat{\mathds{1}})^\dagger(\sqrt{\gamma}\hat{a} + \beta_1\hat{\mathds{1}}) \rangle dt\simeq [\beta_1^2 \hat{\mathds{1}}+ \beta_1 \sqrt{\gamma} \langle (\hat{a}+\hat{a}^\dagger) \rangle] dt \nonumber,\\ 
P_2(dt)&=\langle \hat{K}^\dagger_2 \hat{K}_2\rangle dt=\langle (\sqrt{\eta}\hat{a}^2 + \beta_2\hat{\mathds{1}})^\dagger(\sqrt{\eta}\hat{a}^2 + \beta_2\hat{\mathds{1}}) \rangle dt \simeq \left[ \beta_2^2 \hat{\mathds{1}}+ \beta_2 \sqrt{\eta} (\hat{a}^2+\hat{a}^{\dagger \,2})\right] dt,
\end{align}
where the approximations are valid in the ideal limit $\beta_{1,2}\gg1$.
In this regime, for any time interval, there would occur a huge number of jumps in the total field. 
This would make a naive implementation computationally very demanding, since one should take an extremely small time step.
However, the detected field is almost entirely due to the reference lasers, associated to the operators $\beta_{1,2}\hat{\mathds{1}}$.
Hence, the total jump operators $\hat{K}_{1,2}$ have a very small effect on the resonator state.
In the ideal limit $\beta_{1,2}\to\infty$, the occurrence of an infinite number of jumps is counterbalanced by their infinitesimal effect on the resonator, resulting in an effective diffusive evolution of the cavity state.
The latter, indeed, is found to obey to a stochastic Schr\"odinger equation of the form
\begin{equation}\label{Eq:StocSchHomodyne}
\begin{split}
d\lvert{\psi(t)}\rangle = -\imath\, dt\, \hat{\HH}\, \ket{\psi(t)}
+ \sum_{\nu=1,2} \left \{ 
\left [ \hat{J}_\nu - \frac{{\langle \hat{J}_\nu + \hat{J}_\nu^\dagger \rangle}(t)}{2}\hat{\mathds{1}} \right ] dW_\nu(t) \right.
\\
\left.
- \frac{1}{2} \left [ \hat{J}_\nu^\dagger \hat{J}_\nu
- \langle \hat{J}_\nu + \hat{J}_\nu^\dagger \rangle (t)\hat{J}_\nu  + \frac{{\langle \hat{J}_\nu + \hat{J}_\nu^\dagger \rangle}^2(t)}{4}\hat{\mathds{1}} \right ] dt
\right \}  \ket{\psi(t)},
\end{split}
\end{equation}
where $\hat{J}_{1,2}$ are the resonator jump operators and $dW_{1,2}$ are stochastic Wiener increments of zero expectation value satisfying $dW_\nu(t)dW_\mu(t) = \delta_{\nu\mu}\,dt$ (the detailed derivation can be found e.g. in~\cite{WisemanBOOK}).
Those Wiener processes describe the fluctuation of the homodyne signal.
Using the stochastic Schr\"odinger equation~\eqref{Eq:StocSchHomodyne}, one can simulate the trajectory  by taking a reasonably small $dt$ and generating stochastic Wiener increments at each time step.
Note that Eq.~\eqref{Eq:StocSchHomodyne} does not depend on the values of $\beta_{1,2}$, which are both infinitely large.
In conclusion, the homodyne detection reduces to a continuous diffusive evolution of the wave function.

\section{Intrinsic bimodality in two-photon driven resonators}

As it stems from Eqs.~\eqref{Eq:MixtureCats}~and~\eqref{Eq:MixtureCoherent}, the steady-state density matrix of the system can be cast as the statistical mixture of only two pure states.
This bimodality is an intrinsic property of the Lindblad master equation~\eqref{Eq:Lindblad} and, being an average property of the system, it should somehow appear also on a single experimental realisation.
In other words, the quantum trajectory approach should show a bimodal behaviour.
However, the states between which the system switches, as well as the characteristic time scales, can not be inferred from the form of $\hat{\rho}_{\rm ss}$, and are not manifest in the Lindblad master equation, but depend on the measurement process.

\begin{figure}[t!]
	\begin{center}
		\includegraphics[width=.4\textwidth]{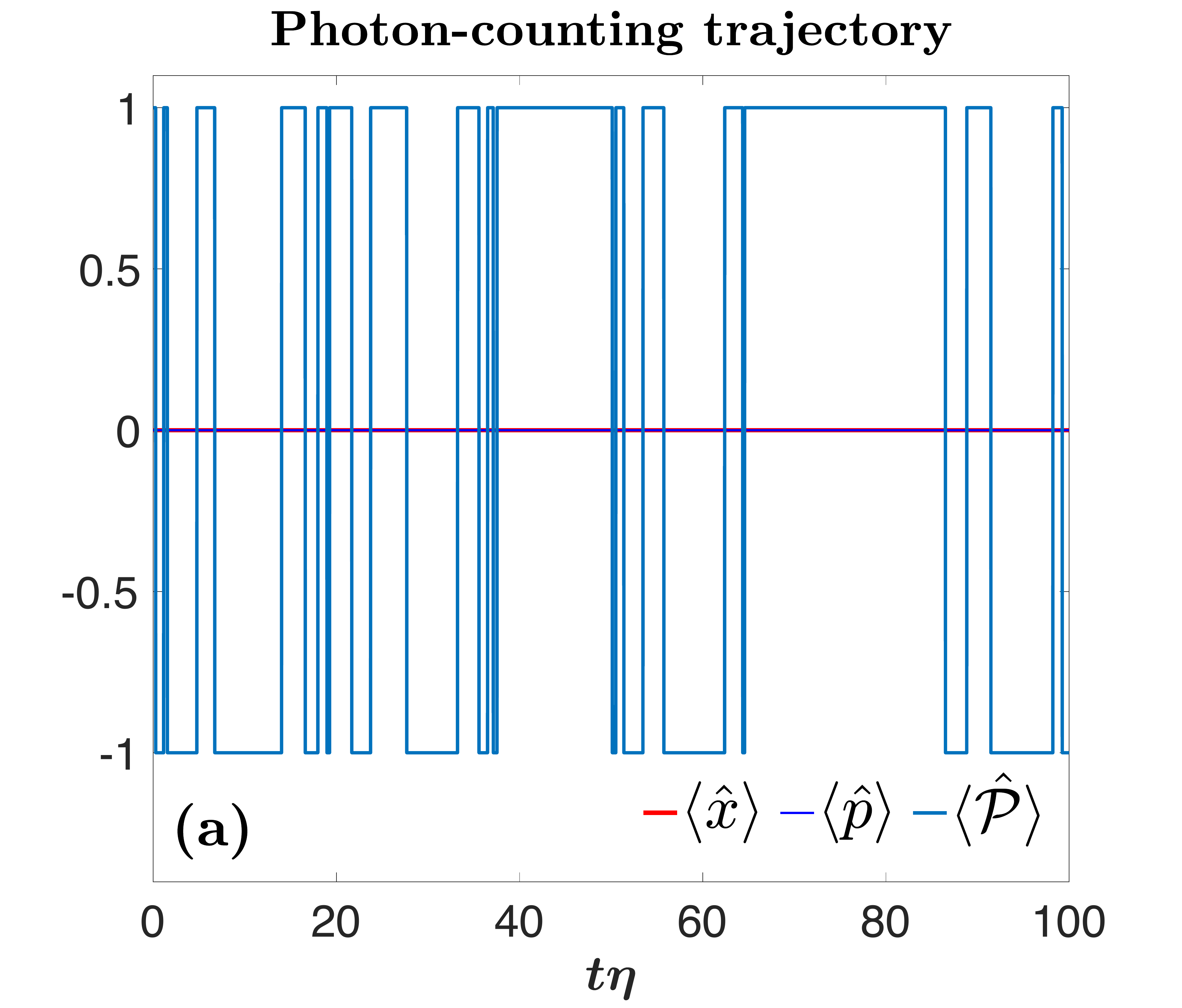}
		\includegraphics[width=.4\textwidth]{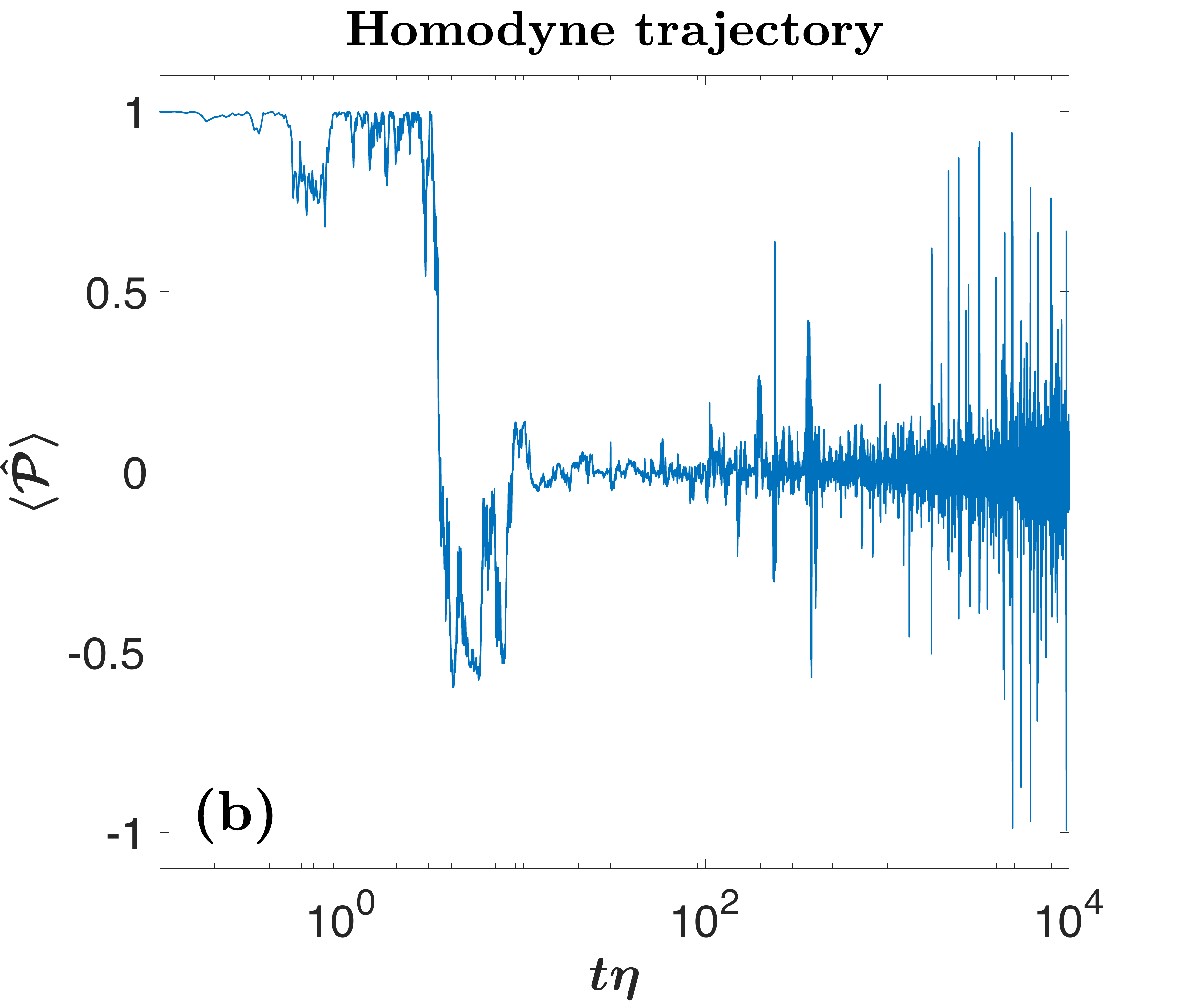}\\
		\includegraphics[width=.9\textwidth, height=0.4\textwidth]{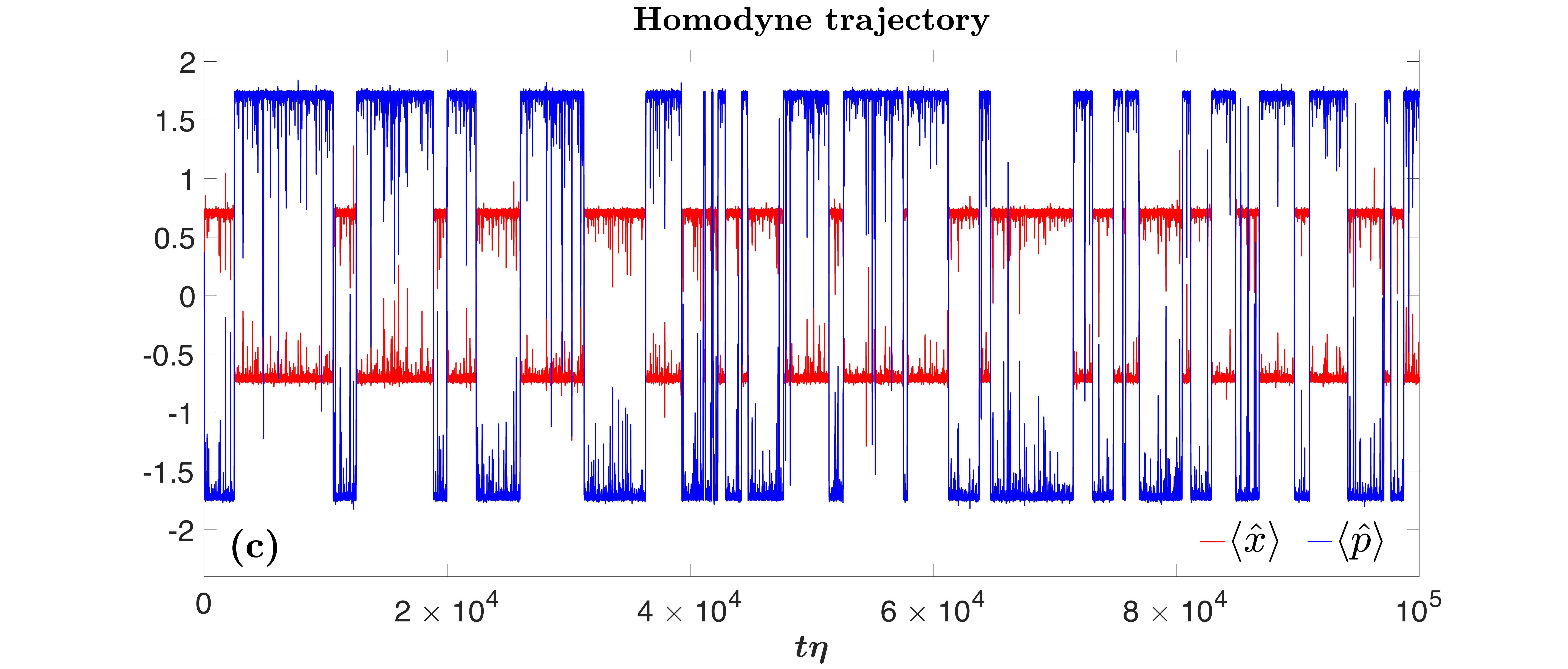}
		\caption{
			Time evolution of $\braket{\hat{x}}$, $\braket{\hat{p}}$, and $\braket{\hat{\mathcal{P}}}$ along single quantum trajectories for the master equation~\eqref{Eq:Lindblad}.
			In~(a) we show the three mean values for a photon-counting trajectory.
			In~(b) and~(c), we plot the same quantities for a homodyne protocol.
			Overall, we set the system parameters to $U=1\eta$, $G=5\eta$, and $\gamma=0.1\eta$.
			Simulations were performed on a truncated Fock basis with $n_{\rm max}=15$, ensuring convergence.
			\label{Fig:TwoPhotonProcesses}}
	\end{center}
\end{figure}

\subsection{Schr\"odinger cats vs coherent states}
\label{Sec:Jumps}
As shown in~\cite{MingantiSciRep16}, the Hamiltonian~\eqref{Eq:Hamiltonian} and the two-photon dissipation tend to stabilize photonic cat states.
On the other hand, the annihilation operator switches from the even (odd) cat to the odd (even) one:  $\hat{a}\ket{\CC^\pm_\alpha} \propto \alpha \ket{\CC^\mp_\alpha}$.
The operator $\hat{J}_1$ thus induces jumps between the two cat states at a rate proportional to $\gamma \braket{\hat{a}^\dagger \hat{a}}$.
This picture is very well captured in the framework of photon-counting trajectories, an example of which is given in 
Fig.~\ref{Fig:TwoPhotonProcesses}(a).
The cat states are, indeed, orthogonal eigenstates of the parity operator $\hat{\mathcal{P}}=e^{\imath \pi \hat{a}^\dagger \hat{a}}$ with eigenvalues $\pm1$.
As we can see, along a single trajectory the state intermittently and randomly switches between the two cat states.
We stress that, instead, the mean values of the field quadratures $\hat{x}=\left(\hat{a}^\dagger+\hat{a}\right)/2$ and $\hat{p}=\imath\left(\hat{a}^\dagger-\hat{a}\right)/2$ are practically zero along the trajectory, as expected for any cat state.
The parity, hence, appears to be the appropriate observable to detect a bimodal behaviour in a photon-counting environment.
Thus, we may interpret $p^\pm$ in Eq.~\eqref{Eq:MixtureCats} as the steady-state probabilities to find the system in one of the two cat states.

The previous analysis seems to point in the direction of privileging Eq.~\eqref{Eq:MixtureCats} over Eq.~\eqref{Eq:MixtureCoherent} as the more truthful picture of the steady state.
This is no more the case if we consider homodyne quantum trajectories.
In Fig.~\ref{Fig:TwoPhotonProcesses}(b), we present (in a log-linear scale) the mean parity $\braket{\hat{\mathcal{P}}}$ along a single homodyne trajectory, taking the vacuum as initial state.
In spite of the ``switching cat'' picture, the parity rapidly approaches zero, and than just fluctuates around this value.
These fluctuations are due to the diffusive nature of Eq.\eqref{Eq:StocSchHomodyne}, which rules the stochastic time evolution of the system wave function under homodyne detection.
The bimodal behaviour, instead, is clear in the time evolution of $\braket{\hat{x}}$ and $\braket{\hat{p}}$, shown in Fig.~\ref{Fig:TwoPhotonProcesses}(c).
This appears compatible with the picture given by Eq.~\eqref{Eq:MixtureCoherent}: at the steady state the system switches between the coherent states $\ket{\pm\alpha}$.
We point out that the phase switches observed for homodyne trajectories have a much smaller rate than parity switches in photon-counting trajectories.
This is a consequence of the metastable nature of the coherent states $\ket{\pm\alpha}$ \cite{LeghtasScience15,MingantiSciRep16}.

Summing up, we have shown that the behaviour of the system along a single quantum trajectory dramatically depends on the measurement protocol adopted.
For photon-counting measurements on the environment, the system switches between the parity-defined cat states appearing in Eq.~\eqref{Eq:MixtureCats}.
Under homodyne detection, the states explored along a single quantum trajectory are the coherent ones in Eq.~\eqref{Eq:MixtureCoherent}.
In other words, one may assign a physical meaning to the probabilities appearing in the mixed-state representation of $\hat{\rho}_{\rm ss}$ only upon specification of the single-trajectory protocol.
However, any possible controversy at the single-trajectory level is washed out by averaging over many of them.

\begin{figure}[t!]
	\begin{center}
		\includegraphics[width=.4\textwidth]{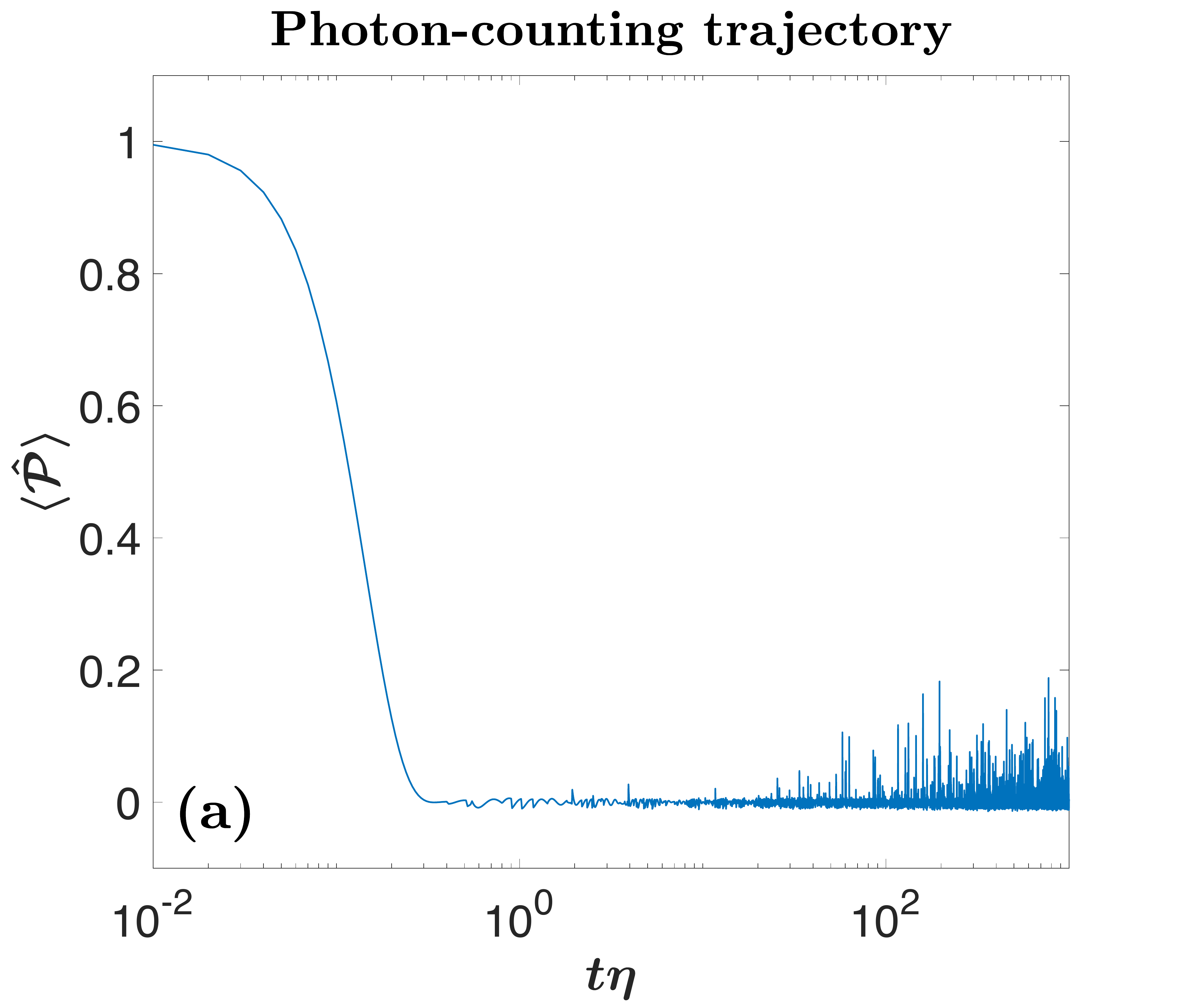}
		\includegraphics[width=.4\textwidth]{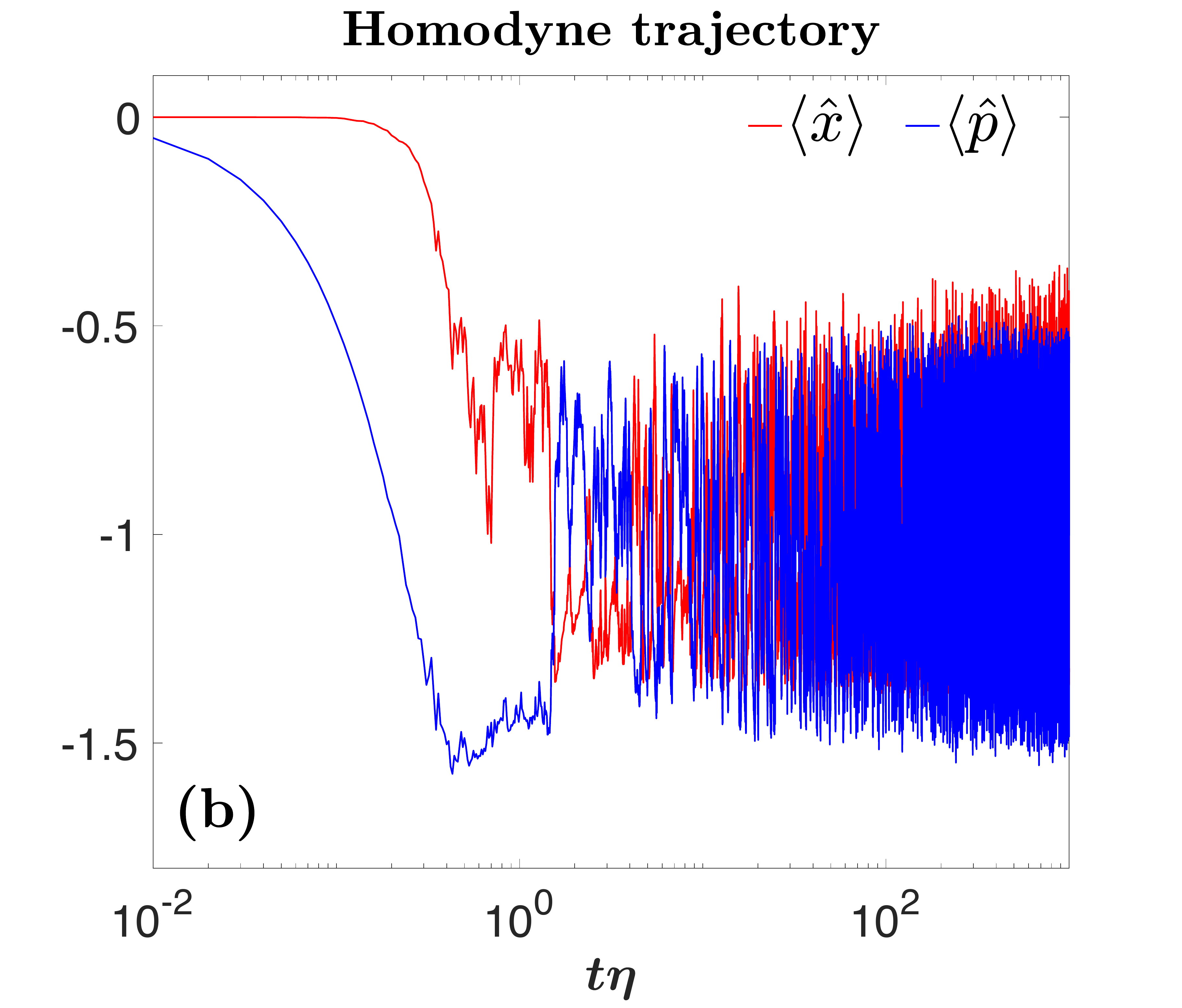}
		\caption{
			Panel~(a): Time evolution of $\braket{\hat{\mathcal{P}}}$ along a single photon-counting trajectory.
			Panel~(b): Time evolution of $\braket{\hat{x}}$ and $\braket{\hat{p}}$ along a single homodyne trajectory.
			Both plots refers to the Lindblad equation~\eqref{Eq:Lindblad} for the one-photon-driving Hamiltonian~\eqref{Eq:HamiltonianF}.
			We set the system parameters to $U=1\eta$, $F=5\eta$, and $\gamma=0.1\eta$ (we stress that here $G=0$).
			Simulations were performed on a truncated Fock basis with $n_{\rm max}=15$, ensuring convergence.
			\label{Fig:OnePhotonProcesses}}
	\end{center}
\end{figure}

\subsection{One-photon driven resonators}

It is legit to question if the abrupt switches observed in the quantum trajectories presented in Fig.~\ref{Fig:TwoPhotonProcesses} are an intrinsic property of the system or is just an effect of the measurement protocol.
To dispel all doubts, we calculated single photon-counting and homodyne trajectories for a resonator subject to a resonant one-photon driving of frequency $\omega_c$.
In the frame rotating at $\omega_c$, the corresponding Hamiltonian reads
\begin{equation}\label{Eq:HamiltonianF}
\hat{\HH}
=\frac{U}{2}\,\hat{a}^\dagger\hat{a}^\dagger\hat{a}\hat{a}
+F \left(\hat{a}^\dagger+\hat{a}\right).
\end{equation}
We stress that, differently from the case discussed above, the steady state of this system is not an equiprobable two-state statistical mixture~\cite{WallsBOOK,DrummondJPA80}.
A photon-counting trajectory for $\braket{\hat{\mathcal{P}}}$ and homodyne trajectories for $\braket{\hat{x}}$ and $\braket{\hat{p}}$ are shown, respectively, in Fig.~\ref{Fig:OnePhotonProcesses}(a) and~(b).
Clearly, the trajectory does not show the same kind of abrupt switches observed in Fig.~\ref{Fig:TwoPhotonProcesses}. 
This proves that the behaviour discussed in Sec.~\ref{Sec:Jumps} is not caused solely by the measurement protocol, but is indeed linked to the bimodal character of the steady state.

\section{Conclusions and Perspectives}\label{Sec:Conc}

In this article, we have studied the quantum many-body behaviour of interacting photons in a nonlinear resonator subject to engineered two-photon processes.
The objective has been to point out and characterize the bimodal nature of the steady state, which can be seen, equivalently, as the statistical mixture of photonic Schr\"odinger cat states [Eq.~\eqref{Eq:MixtureCats}] or of coherent states with same amplitude and opposite phases [Eq.~\eqref{Eq:MixtureCoherent}].
Resorting to the Montecarlo wave function method, we have shown that, along a single quantum trajectory, the bimodal nature manifests through abrupt switches along the time evolution of some observables. However, depending on the model adopted for the detection protocol, these switches appear in different observables.
The two protocols described in this work seem to privilege  one of the two representations~\eqref{Eq:MixtureCats} and~\eqref{Eq:MixtureCoherent}, creating an apparent contradiction.
However, this issue is lifted when looking at the average behaviour, which is an actual implementation of the Lindblad master equation~\eqref{Eq:Lindblad}.
Finally, we have also studied the quantum trajectories for a one-photon-driven resonator in a regime where its steady state is not bimodal.
The absence of abrupt switches in parity or quadratures proves that the ones observed in Fig.~\ref{Fig:TwoPhotonProcesses} are not artefacts of the quantum trajectory approach, but a feature linked to the steady-state bimodality.

The results presented in this work shed light on the physical interpretation of quantum trajectories issued from Montecarlo wave function algorithms, which are already widely applied in the study of out-of-equilibrium systems.
Investigating the onset of bimodality in driven-dissipative resonators is particularly interesting since it has been related with the occurrence of a first-order dissipative phase transition~\cite{BartoloPRA16}.
Furthermore, the components of the mixed steady state~\eqref{Eq:MixtureCats} or~\eqref{Eq:MixtureCoherent} can be used as (quasi-)orthogonal states in quantum computation~\cite{RalphPRA03,GilchristJOB04,MirrahimiNJP14,PuriarXiv16}.
To exploit the two-photon driven resonator in this context, one can envision a feedback mechanism which unbalances the steady-state mixture in favour of one of the two components~\cite{MingantiSciRep16}.
Since a feedback relies on measurements, the quantum-trajectory analysis is extremely helpful in the design of efficient and practical feedback protocols.

\smallskip
We acknowledge fruitful discussions with W. Casteels and the support from ERC through the Consolidator Grant “CORPHO” No.616233.


\end{document}